\documentclass[12pt]{article}
\usepackage[T2A]{fontenc}
\usepackage[utf8]{inputenc}
\usepackage[english]{babel}
\usepackage{graphicx}
\usepackage{caption}
\usepackage{cyr}
\usepackage{epsf}
\usepackage{pazh}
\usepackage{xcolor}
\usepackage{amsmath,amssymb}

\usepackage{longtable}

\tightenlines
\usepackage{physics}
\usepackage{textcomp}

\def\a{$^{\mbox{\small a}}$}
\def\b{$^{\mbox{\small b}}$}
\def\c{$^{\mbox{\small c}}$}
\def\d{$^{\mbox{\small d}}$}
\def\e{$^{\mbox{\small e}}$}
\def\etal{{et~al.}}
\sloppypar

\begin{document}
\baselineskip 21pt


\title{\bf Extended Emission of Cosmic Gamma-Ray Bursts Detected in the
SPI-ACS/INTEGRAL Experiment}

\author{\bf \hspace{-1.3cm}\copyright\, 2020 г. \ \
G. Yu. Mozgunov\affilmark{1,2*}, P. Yu. Minaev\affilmark{1,2} and A. S. Pozanenko\affilmark{1,2}}

\affil{
{\it Space Research Institute, Russian Academy of Sciences, Profsoyuznaya ul. 84/32, Moscow, 117997 Russia}$^1$\\
{\it Moscow Institute of Physics and Technology, Institutskii per. 9, Dolgoprudnyi, Moscow obl., 141701 Russia}$^2$\\
}
\vspace{2mm}

\sloppypar
\vspace{2mm}
\noindent
{\bf Abstract --} We have carried out a systematic analysis of the gamma-ray bursts' (GRBs) light curves detected in the SPI-ACS experiment onboard the INTEGRAL observatory aimed to search extended emission. The emission occasionally recorded after the prompt active phase of a GRB in the form of an emission that is longer than the active phase and less intense is called the extended one. Out of the 739 brightest GRBs recorded from 2002 to 2017, extended emission has been detected in $\sim20\%$ of the individual light curves; its maximum duration reaches $\sim 10000$ s. Two different types of extended emission have been revealed. One of them is an additional component of the light curve and is described by a power law (PL) with an index $\alpha \sim -1$ close to the PL index of the afterglow in the optical and X-ray bands. The second type can be described by a steeper PL decay of the light curve typical of the active burst phase. Extended emission has also been found in the combined light curve of long GRBs in the individual curves of which no extended emission has been detected. The PL index of the extended emission in the combined light curve is $\alpha \sim -2.4$. It is most likely associated with the superposition of light curves at the active phase; its total duration is $\sim 800$ s.



\vfill
\noindent\rule{8cm}{1pt}\\
{$^*$$<$georgiy99@bk.ru$>$}

\clearpage

\section*{INTRODUCTION}
\noindent

Gamma-ray bursts (GRBs) were discovered as powerful bursts of gamma-ray radiation in 1967 (Klebesadel et al. 1973). Since then the nature of the phenomenon has remained one of the interesting puzzles in astrophysics. Bimodality in the GRB duration distribution was first detected in Konus experiments (Mazets et al. 1981). It was confirmed by Kouveliotou et al. (1993) based on extensive statistical data from the BATSE experiment and, subsequently, in the data of many other experiments (see, e.g., Minaev et al. 2010a, 2010b, 2012, 2014; Minaev and Pozanenko 2017, 2020). This suggested the existence of two classes of bursts: short/hard ones (less than 2 s in duration) associated with neutron star mergers (Paczyn´ski 1986; Abbott et al. 2017) and long/soft ones (more than 2 s in duration) associated with the collapse of massive stars (Woosley 1993). Two phases of the emission process are distinguished. The active phase is characterized by the operation of the central engine, during which active energy release occurs and a jet is formed. During the active phase a prompt gamma-ray emission (the GRB itself) is produced. The passive phase follows after the termination of the central engine operation the jet propagates in external medium accompanied by adiabatic cooling and radiation in the X-ray, optical, and radio bands (see, e.g., Meszaros and Rees 1992) called the afterglow.

The emission occasionally recorded after the prompt active phase of a GRB in the form of an emission that is more prolonged than the active phase and significantly less intense is called the extended one. This emission was detected, for example, in the data from the SIGMA/Granat experiment in the individual and combined light curves (Burenin et al. 1999, 2000; Tkachenko et al. 2000; Burenin 2000). A statistical study of the averaged light curves for GRBs based on data from the BATSE experiment revealed extended emission with a duration up to 100 s in short bursts and 1000 s in long ones (Lazzati et al. 2001; Connaughton 2002). Extended emission in the range 10–100 keV with a duration of 100 s was found in the averaged light curve of short GRBs based on Konus/Wind data (Frederiks et al. 2004). Extended emission from short GRBs was also found in some individual light curves in the BAT/Swift (Kaneko et al. 2015; Norris et al. 2011), GBM/Fermi (Kaneko et al. 2015), and Konus/Wind (Mazets et al. 2002) experiments, with it having been spectrally softer than the main episode. In the harder energy range > 80 keV in the SPI-ACS/INTEGRAL experiment (Minaev et al. 2010а) extended emission was also detected from short bursts both in the combined light curve with a duration of 125 s and in the individual light curves of several bursts.

Two types of extended emission probably exist in short GRBs. This is confirmed by the fact that the extended emission continues the main phase by a monotonic decay in some cases (see, e.g., Svinkin et al. 2016; Minaev et al. 2017) and has a complex structure consisting of several pulses in other cases (for example, in the case of GRB 060614 Gehrels et al. (2006)). Extended emission from long bursts was found in the individual light curves (Giblin et al. 2000; Ronchi et al. 2019), where it was described by a power law model and the spectrum was typical of the synchrotron radiation from a jet propagating in external medium, i.e., the afterglow. No systematic study of the extended emission from long bursts has been carried out since 2002.

Various models have been proposed to explain the nature of the extended emission. It can be the spectrally hard part of the afterglow (Burenin 2000; Lazzati et al. 2001). In the model proposed, for example, by Metzger et al. (2008) the extended emission is maintained by the rotational energy of a magnetar. In the two-jet model the extended emission and the emission at the active phase are maintained by different mechanisms (Barkov and Pozanenko 2011). The extended gamma-ray emission can also be associated with charged particle acceleration processes at the shock front (Warren et al. 2018). The interaction of the jet and the breakout shell, the cocoon radiation (Gottlieb et al. 2018; Pozanenko et al. 2018), may also be considered as a mechanism of the extended emission.

In bursts with extended emission the well-known duration parameter $T_{90}$ can reach thousands of seconds. Occasionally, such bursts can be confused with ultra-long ones, i.e., GRBs with a duration of the active phase of 1000 s or more (Gendre et al. 2013). However, the light curve of such bursts most often consists of several emission episodes separated by quiescence periods, which is typical of the active phase and not the extended emission.

The nature of the extended emission remains unclarified. No systematic study of the phenomenology of long bursts based on which it would be possible to choose an optimal model of the extended emission has been carried out so far. In this paper we carry out a systematic study of the extended emission of GRBs using data from the anticoincidence shield (ACS) of the SPI spectrometer onboard the INTEGRAL observatory in the energy range >80 keV.

\section*{DATA SELECTION AND PROCESSING}
\noindent
\subsection*{SPI-ACS INTEGRAL}
\noindent

SPI-ACS is the anticoincidence shield of the germanium SPI detector onboard the INTEGRAL space observatory. Ninety one BGO (bismuth germanate) scintillators surrounding the SPI telescope are used as detectors. Two photomultiplier tubes (PMTs) are coupled with each BGO crystal and the counts from all PMTs are recorded in a single energy channel. The lower and upper thresholds of the channel are 80 keV and	10 MeV, respectively. The SPI-ACS experiment can record photons from all directions, but the direction coincident with the SPI field of view, $16^{o}$ is least sensitive. The detector time resolution is 50 ms (von Kienlin et al. 2003). The data used here are publicly accessible (https://www.isdc.unige.ch/savchenk/spiacs-online/spiacs-ipnlc.pl).

The INTEGRAL space observatory is in a highly elliptical orbit with an initial orbital period of 72 h and an apogee 153 000 km. Such an orbit provides background stability on long time scales compared to near-Earth spacecraft. An analysis of the background on the near-Earth Fermi spacecraft can be found, for example, in Biltzinger et al. (2020).
The background in the GBM/Fermi experiment can change several-fold on time scales of only a few hundred seconds. All of this makes it diffcult to study a prolonged and weak signal, such as the extended emission. Minaev et al. (2010a) and Bisnovatyi-Kogan et al. (2011) showed that the SPI-ACS background level changes by no more than 0.3\% on time scales up to a thousand seconds. Despite the limitations of the SPI-ACS experiment, namely the recording of photons in a single energy channels with a time resolution of 50 ms, a stable background on long time scales makes it an effective tool for studying the extended emission.

\subsection*{DATA PROCESSING}
\noindent

To produce a sample of GRBs, we used the master list by Hurley (2008) (http://www.ssl.berkeley.edu/ipn3/masterli.txt). It is a compilation of the data from a large number of space experiments recording GRBs, including SPI-ACS, collected by K. Hurley and maintained from 1990 until now. According to the data from the master list, 4720 GRBs were detected from November 2002 to November 2017. Our sample was produced from them using the following criteria applied to the SPI-ACS data: (1) a burst detection significance more than $25\sigma$ on a time scale of 1 s, which is equivalent to a peak ﬂux $\sim10^4$ counts per second; (2) the absence of telemetry gaps in the interval (0, 100) s relative to the trigger time. It is well known from early studies (Connaughton 2002) that the duration of the extended emission is $10^2 - 10^3$ s. Therefore, the chosen interval allows the undistorted extended emission or at least its onset to be distinguished. A total of 786 events satisfied these criteria.

The data in the time interval $[L; R]$ relative to $T_0$, where $L, R = 15 000$ s, are used in the search for extended emission. The original time resolution (bin duration) is 0.05 s. The time intervals to fit the background are chosen as follows: the left one is $[ L; 0.1\cdot L]$ and the right one is $[0.4\cdot R; R]$. The intervals are not symmetric; the left end of the right interval is offset from $T_0$ farther than the right end of the left one. This choice of intervals can provide a more accurate separation of the extended emission that is expected after the GRB itself. Then, the background signal is fitted by two models: linear and cubic polynomials. The best background model is determined from the reduced value of the functional $\chi^2/d.o.f.$ If the value of the functional $\chi^2/d.o.f. ~\gg 1$ in both cases, then $L$ and $R$ are reduced and the and the process is repeated. After determining the best model in the maximum possible background interval, the background model is subtracted from the original light curve.

The duration of the background intervals $R + L$ is varied from $10^2$ to $10^4$ s. $L, R \sim 10^3$ s are used most commonly. It turned out that an optimal background model for most bursts is a cubic polynomial (51\% of all the investigated ones), in 336 cases (43\%) the background is described by a linear model, and in 47 cases (6\%) the background has a more complex form and cannot be satisfactorily described by these models.

The sample mean (M) and sample dispersion (D) are calculated for each burst in the intervals of the background fit. We detected a deviation from the Poisson distribution; the dispersion is systematically higher than the mean, confirming the results obtained by Rau et al. (2005), Ryde et al. (2003), and Minaev et al. (2010a). The ratio of these quantities $k=\frac{D}{M}$ changes in the range  $1.18<k<1.76$. The $1\sigma$ significance level of the signal above the background $B$ is determined from the formula $\sqrt{k \cross B}$.

\begin{figure}[h]
\hspace{-2cm}
\center{\includegraphics[width=1\linewidth]{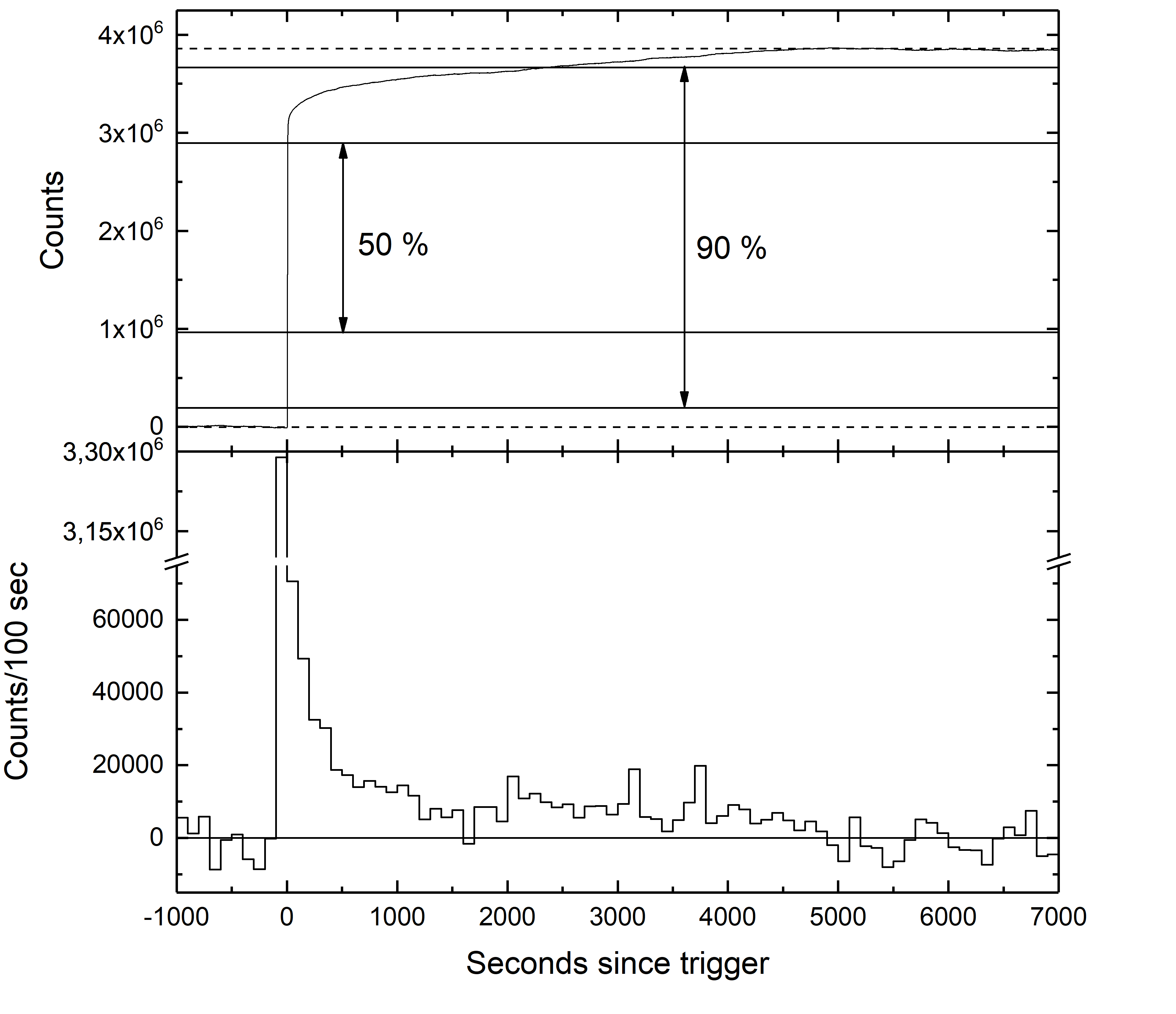}}
\caption{\rm Integral curve (top) and light curve of GRB 021206 (bottom); the bin duration is 100 s. The time relative to the trigger is along the horizontal axis. The dashed lines on the integral curve mark the 0 and 100\% levels of counts. The solid lines indicate the levels of accumulated counts that are 5, 25, 75, and 95\% of the maximum level.}
\label{T90}
\end{figure}

The next step after the background subtraction is to calculate the duration parameter $T_{90}$ (the time in which 90\% of the counts of the event arrive). The calculation algorithm is described in detail in Koshut et al. (1996). Fig.~\ref{T90} presents the scheme of calculating $T_{90}$ for GRB 021206. Here we will brieﬂy describe this algorithm. To determine the parameter $T_{90}$, it is necessary to construct the integrated light curve. Then, we need to determine the levels corresponding to 5\% and 95\% of the total number of counts in the event in the integrated light curve. Thereafter, the corresponding times $T_{5\% }$ and $T_{95\% }$ are determined. The difference of these times is the duration parameter $T_{90}$. The duration parameter $T_{50}$ can also be calculated in a similar way. Table 1 gives the values of $T_{90}$ calculated in our paper and, where possible, we provide the values of $T_{90}$ obtained in other experiments (GBM/Fermi, BAT/Swift, RHESSI, Konus/Wind, HETE-2). In most cases, there are no significant differences in $T_{90}$. However, the duration parameter $T_{90}$ calculated in our paper for bursts with extended emission exceeds considerably the parameter $T_{90}$ in other experiments,including even the SPI-ACS catalog (Rau et al. 2005). These discrepancies are probably explained precisely by the choice of intervals to fit the background. When choosing an interval close to the main burst phase, part of the interval with extended emission will be used to determine the background and, consequently, part of extended emission will not be included in $T_{90}$ and parameter value will be inderestimated. As an example, let us again consider GRB 021206 process is repeated. After determining the best model in the maximum possible background interval, the (Fig. 1), for which $\frac{T_{90,SPI-ACS}}{T_{90,RHESSI}}\sim450$. When the light curve is analyzed in detail, it turns out that the burst has a significant extended emission representing a light curve with a power law decay and a duration $10^4$ s. Apart from the extended emission, naturally, there is a prompt main phase (active phase) of the burst with a duration of only $\sim10$ s. If we include the extended emission in the background interval, then only the prompt burst phase with a duration close to $T_{90}$ from the RHESSI catalog will remain after the background subtraction.

During the main phase a large ﬂux of counts is concentrated in a short (compared to the duration of the extended emission) time interval. This behavior on the integrated curve is represented by the nearly ver- tical segment. The extended emission is represented by the slow rise after the main burst phase (Fig.~\ref{T90}, the integrated curve, the time interval from $\sim200$ s to $\sim4000$ s). The 25\% and 75\% levels needed to calculate $T_{50}$ cross only the vertical segment of the curve and, therefore, $T_{50}$ characterizes the duration of only the active phase. The 95\% ﬂux level crosses the ﬂat part of the curve and, hence, the duration $T_{90}$ does includes part of the interval of the extended emission. The inﬂuence of the extended emission on the duration parameters of GRBs was also studied by Burenin (2000) and Burenin et al. (2000).

\subsection*{SEARCHING FOR EXTENDED EMISSION}
\noindent

At the initial time resolution (50 ms) the extended emission has a low significance above the background $\lesssim 1\sigma$.. The following light curve processing algorithm is used to increase the statistical significance of the sought signal.

\begin{figure}[h]
\hspace{-2cm}
\center{\includegraphics[width=1\linewidth]{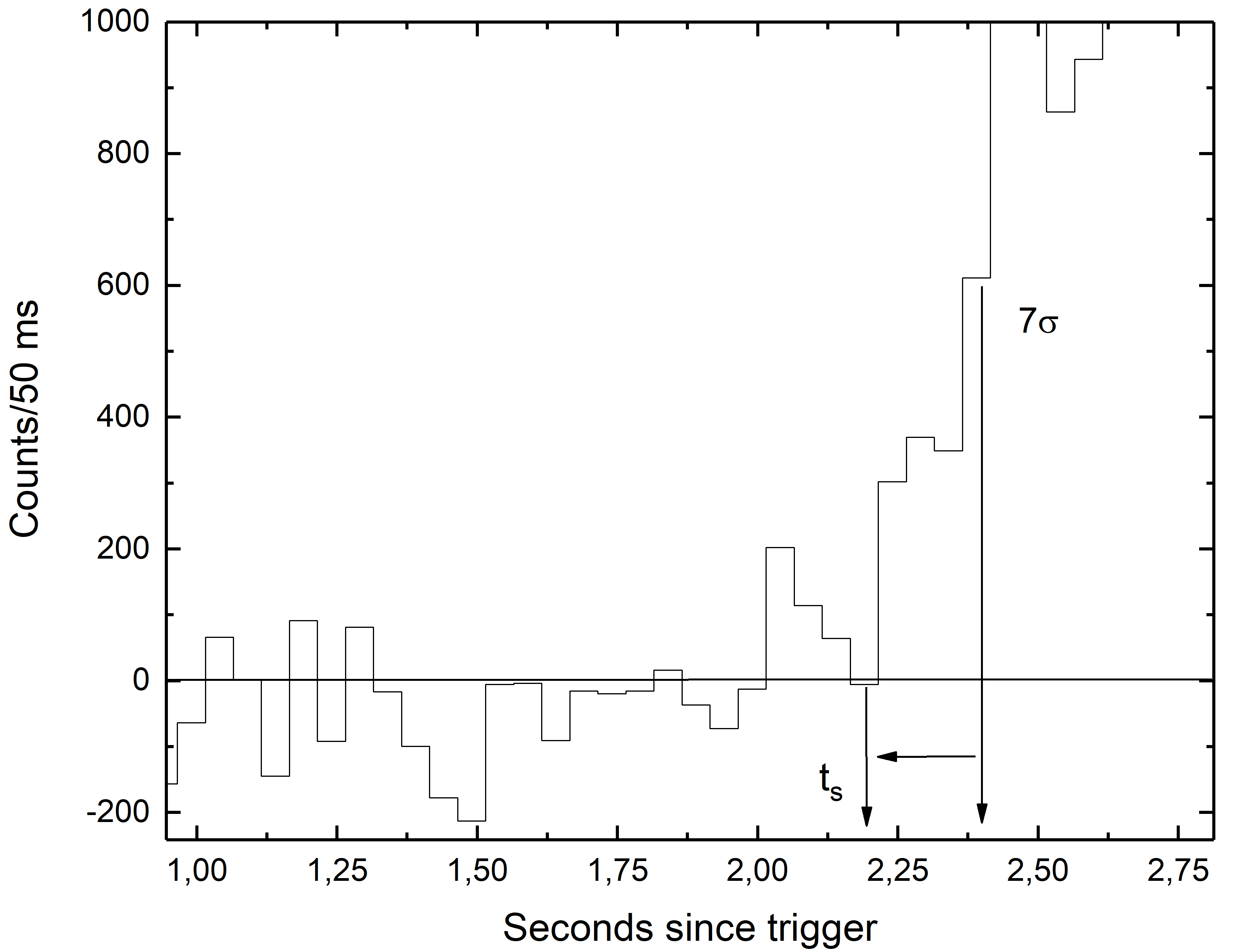}}
\caption{\rm Light curve of GRB 021206 after the background subtraction in the interval [1.00; 2.75] s.}
\label{t_s_calcul}
\end{figure}

The first step is to determine the GRB onset time. The calculation scheme is presented in Fig.~\ref{t_s_calcul}. The first (in time) bin in the original time resolution, in which the significance of the signal from the GRB above the background reaches $7\sigma$, is used as the initial approximation. Then, the closest time preceding this bin at which the ﬂux from the GRB becomes zero is determined -- this is the sought for burst start time $t_s$.

\begin{figure}[h]
\hspace{-2cm}
\center{\includegraphics[width=1\linewidth]{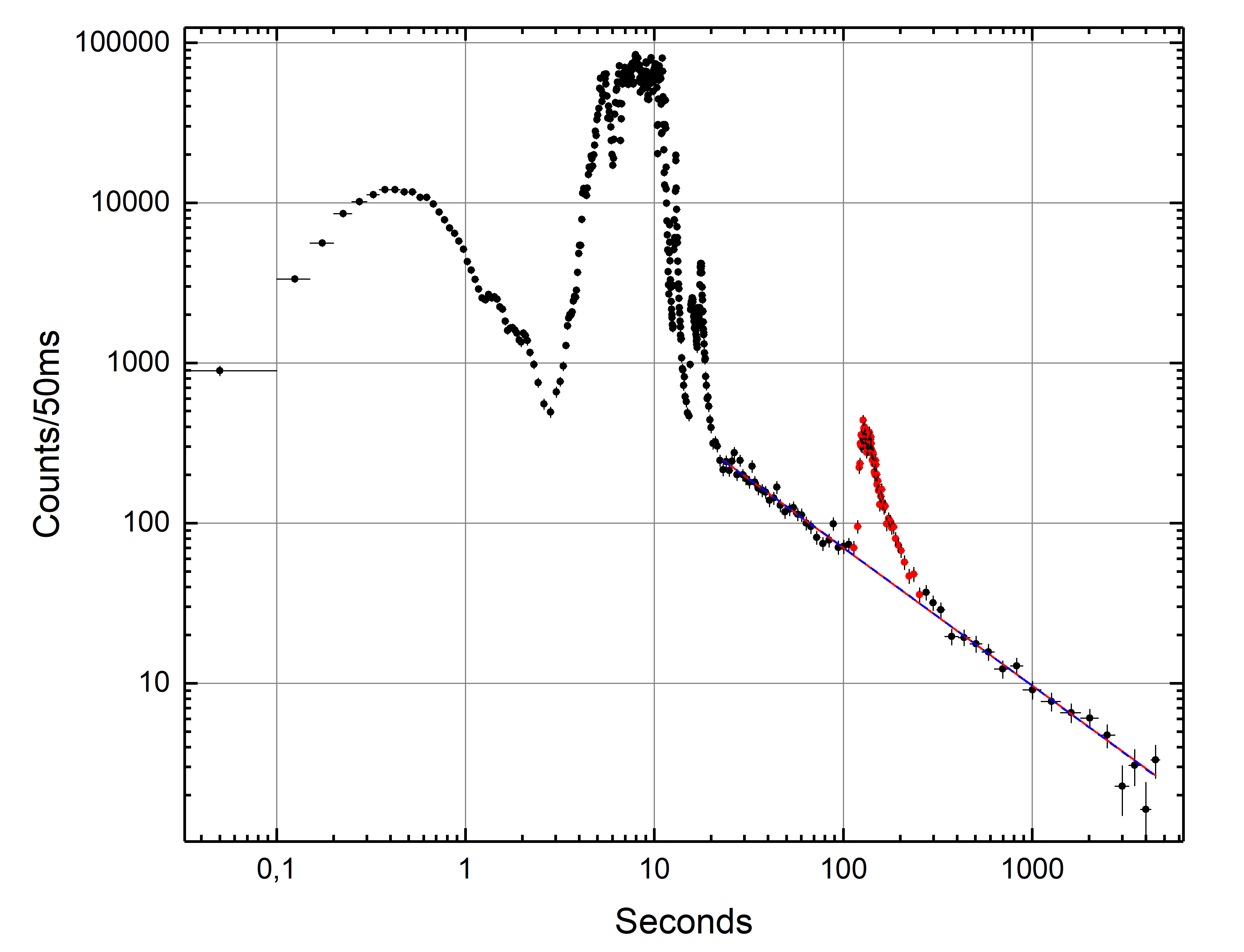}}
\caption{\rm Light curve of GRB 130427. The time relative to the burst start time $t_s$ is along the horizontal axis. The number of counts in 50 ms is along the vertical axis. The bins in the interval [100; 250] s are not involved in the fit. The red and blue curves represent the PL and biased PL fits, respectively.}
\label{extended130427}
\end{figure}

The next step is logarithmic binning. The bins are combined in the original time resolution, beginning from $t_s$, until a certain statistical significance level $\sigma_T$ is reached. The value of $\sigma_T$ depends on the accumulated duration as $\sigma_T=A-B\log_{10}T$, where $T$ is the total duration of the accumulated bins, while $A$ and $B$ are the model parameters chosen individually for each burst. Their typical values are $A\sim 8,B\sim 1.4$. Assuming a power law behavior of the light curve for the extended emission, logarithmic bins allow the $\chi^2$ statistic to be used in the fit, because bins with a large number of counts needed to use the criterion are formed. An example of the light curve with logarithmic binning is presented in Fig.~\ref{extended130427}.

The logarithmic binning is followed by the extended emission fitting. A power law (PL) $C=A\cdot t^{-\alpha}$ is used for this purpose. However, the light-curve shape is not described by PL in some cases. A second model with an additional degree of freedom, the onset time of the extended emission $t_{EE}$, is introduced for these cases: $C=A\cdot (t-t_{EE})^{-\alpha}$ (biased PL). The Bayesian information criterion $BIC=\chi^2+kln(N)$ (Liddle 2007), where $k$ is the number of parameters, $N$ is the number of data points used in the fit, and $\chi^2$ is the value of the functional, is used to determine the best model.	If the value of the criterion for a complex model is smaller than that for a simple one, then the complex model is preferable. For some bursts in the biased PL model $t_{EE}$ takes negative values and, hence, the extended emission begins earlier than ts. The start time of the extended emission is unknown. In this paper we assume that it cannot begin earlier than the burst start time $t_s$. Therefore, $t_{EE}$ cannot take negative values and the biased PL model in these cases cannot be deemed preferable.

As the right boundary of the fitting interval we choose the last significant bin. For the beginning of the fitting procedure we take three successive bins preceding the last one; using these four bins, including the last one (the minimum number of data points for the fitting by a function with three parameters), we perform the fitting by PL and biased PL. Thereafter, the left boundary of the fitting interval is shifted by one bin to the burst start time and the fitting is repeated. The process continues until the deviation of the count rate in the last added bin from the one obtained after the model fitting begins to exceed $3\sigma$. This suggests that the observed count rate in the last added bin is no longer described by the model with a single power law.

\begin{figure}[h]
\hspace{-2cm}
\center{\includegraphics[width=1\linewidth]{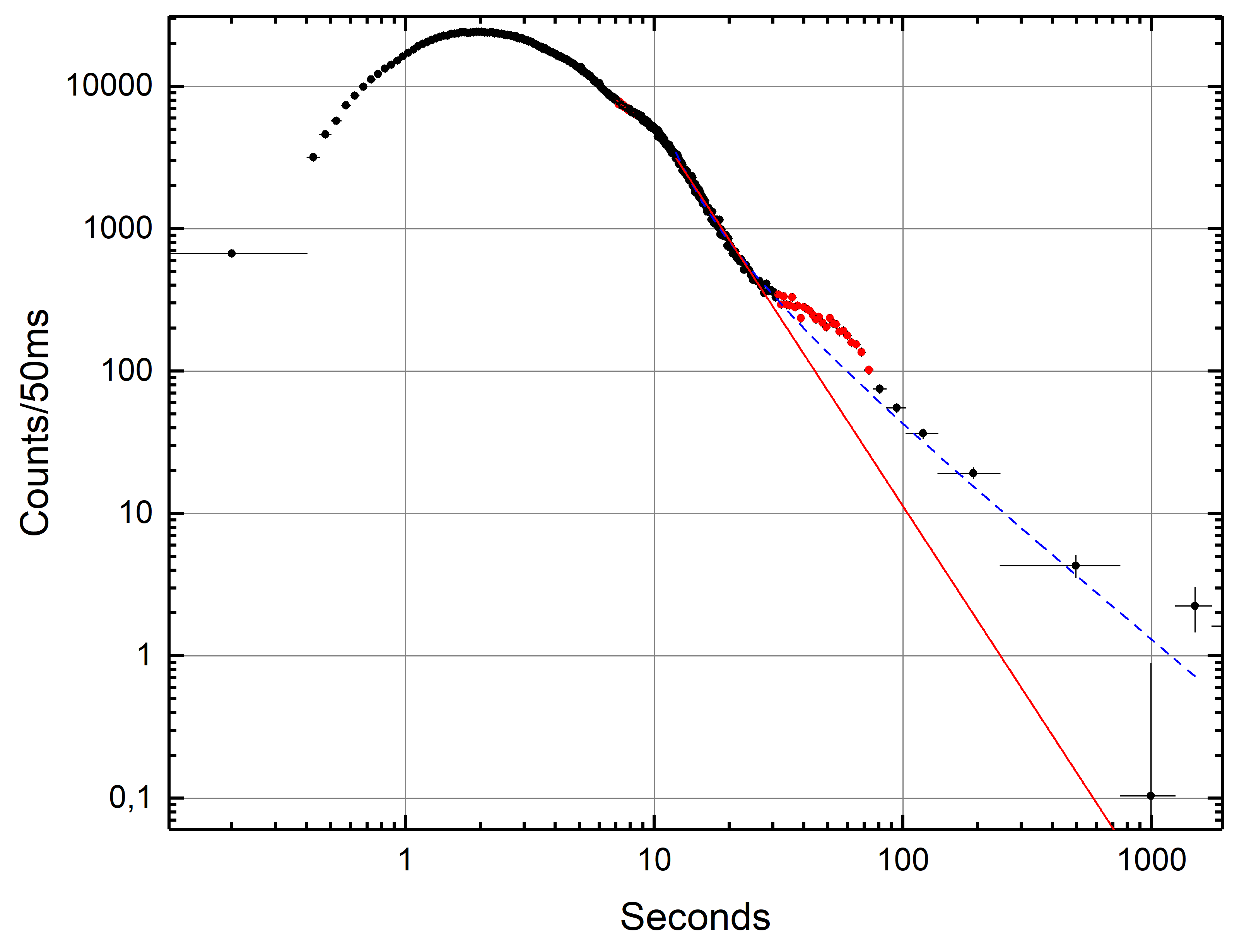}}
\caption{\rm Light curve of GRB 041212. The time relative to the burst start time $t_s$ is along the horizontal axis. The number of counts in 50 ms is along the vertical axis. The red bins are not involved in the fit. The red and blue curves represent the PL and biased PL fits, respectively.}
\label{extended041212}
\end{figure}

Two variants are possible when choosing the fitting interval. The first variant: it is impossible to determine the end time of the active phase domination. In these cases, the PL model describes the burst from the light curve peak to the last significant bin (Fig.~\ref{extended041212}). Hence, the light curve of the active phase is described by PL; such bursts belong to type II extended emission. The second variant: the extended emission is not fitted by a single PL model. In these cases, the last bins cease to be described by the model when the left edge of the fitting interval moves away from the right one. This means that the model already begins to describe part of the active phase, while the extended emission is an additional component (type I extended emission). If the number of bins for the extended emission (less than four) is not enough for PL fitting, then we calculate only the statistical significance of the extended emission.

Consider the light curve of GRB 130427 in Fig.~\ref{extended130427}. The last significant bin is at $\sim5000$ s; it is also the end of the fitting interval. The left edge of the interval is at $\sim20$ s; at this time the second component of the signal, i.e., the extended emission, begins to dominate and a break is present in the light curve. The burst with such a feature in the light curve belongs to type I extended emission. A rise in ﬂux deviating from PL is also observed in the time interval 100–250 s. This rise may not refer to the extended emission and, therefore, is not involved in the fit. The bursts with these features in the light curve are additionally marked by index <<b>> and otherwise index <<a>>.

The results of our search for extended emission are presented in Table 2. It contains 151 GRBs with a significant ($>3\sigma$) extended emission and gives the best model for its description and the model parameters $A,\alpha$ and, for biased PL, $t_{EE}$. Only the significance of the extended emission is presented for 40 bursts.

\subsection*{COMBINING OF LIGHT CURVES}
\noindent

Here we want to test the hypothesis that the ex- tended emission is a common property of all bursts, but in some events it is not detected because of its low significance. If this extended emission is assumed to be actually present in all GRBs, then the statistical significance can be increased by combining the light curves. Indeed, if identical light curves with the same background level are combined (stacked), then the statistical significance of the useful signal will increase by a factor of $\sqrt{N}$, where $N$ is the number of bursts in the sum. GRBs without extended emission in their individual light curves are used to construct the combined light curve.

As has already been discussed in the Introduction, the short and long bursts are different in nature and the extended emission mechanisms can be different. Therefore, the two classes of bursts should be combined separately. Obviously, the distribution in duration parameter $T_{90}$ constructed from the data of this paper is distorted. The burst selection algorithm on a time scale of 1 s rejects the very short bursts. A different selection criterion was used in Minaev et al. (2010b). The boundary value of $T_{90}$ is$ 0.7$ s and we will use it to separate the two classes of events. The separation of the classes of short and long bursts using the $0.7$ s boundary is more justified for SPI-ACS, i.e., for an energy range above $\sim80$ keV.

\begin{figure}[h]
\hspace{-2cm}
\center{\includegraphics[width=1\linewidth]{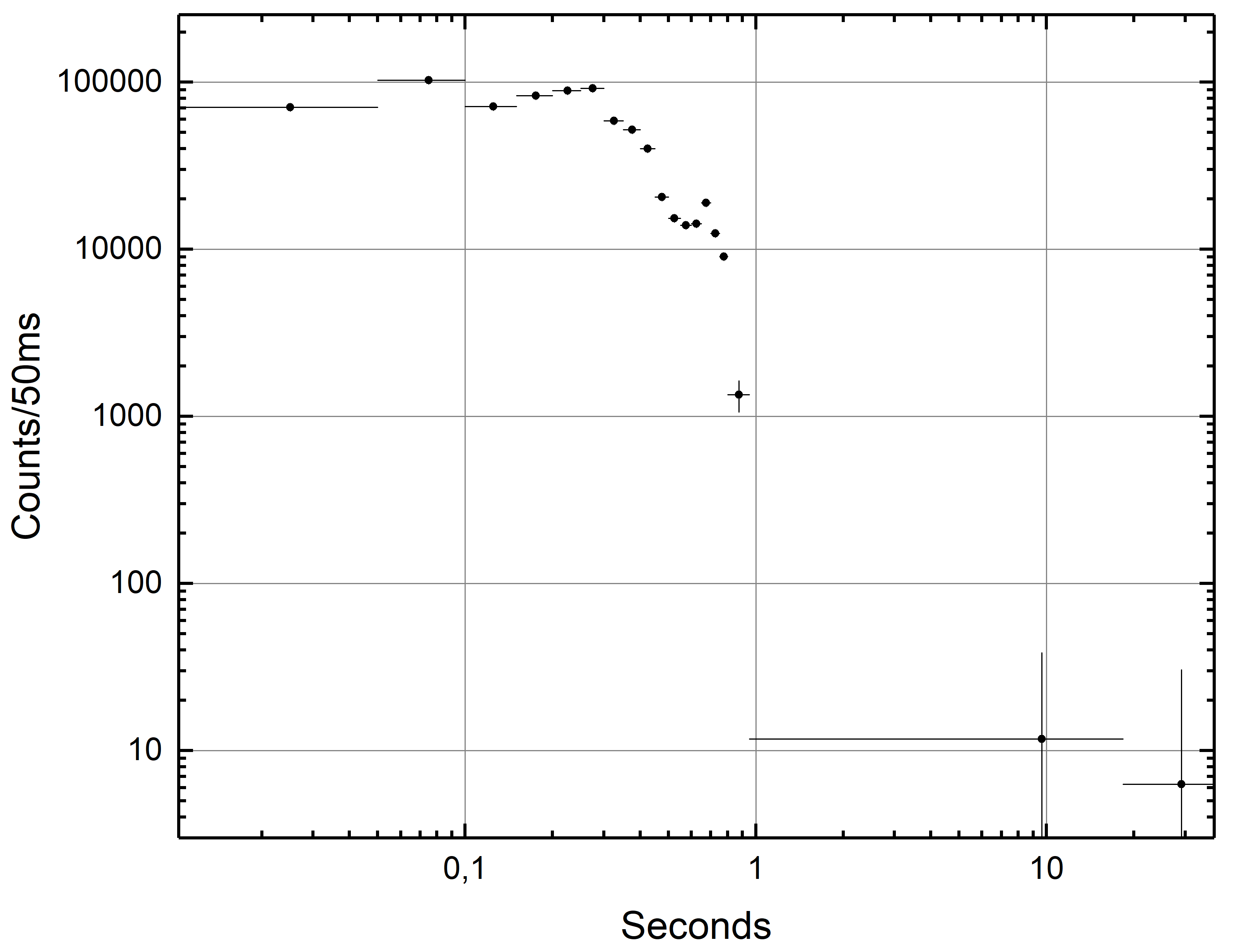}}
\caption{\rm Combined light curve of 40 short GRBs without extended emission in the individual light curves.}
\label{short_sum_log}
\end{figure}

The combined light curve of short bursts was constructed in the interval $[-500;500]$ s. Figure~\ref{short_sum_log} presents the combined light curve of 40 short bursts in a logarithmic scale. The active phase ends approximately at 1 s; no significant emission was detected after it. This allows an upper limit on the ﬂuence of the extended emission to be determined in an interval of 125 s after the end of the main phase. The SPI- ACS calibration (Vigano and Mereghetti 2009) to convert the detector counts to energy units is used for this purpose. For a normal angle between the source and the X axis of the INTEGRAL observatory one SPI-ACS count corresponds to $10^{-10}$~erg~cm$^{-2}$. Taking into account the average background level and the coeffcient $k = 1.35$, we find the ﬂuence in the extended emission to be $\sim2\cross10^{-6}$~erg~cm$^{-2}$.

\begin{figure}[h]
\hspace{-2cm}
\center{\includegraphics[width=1\linewidth]{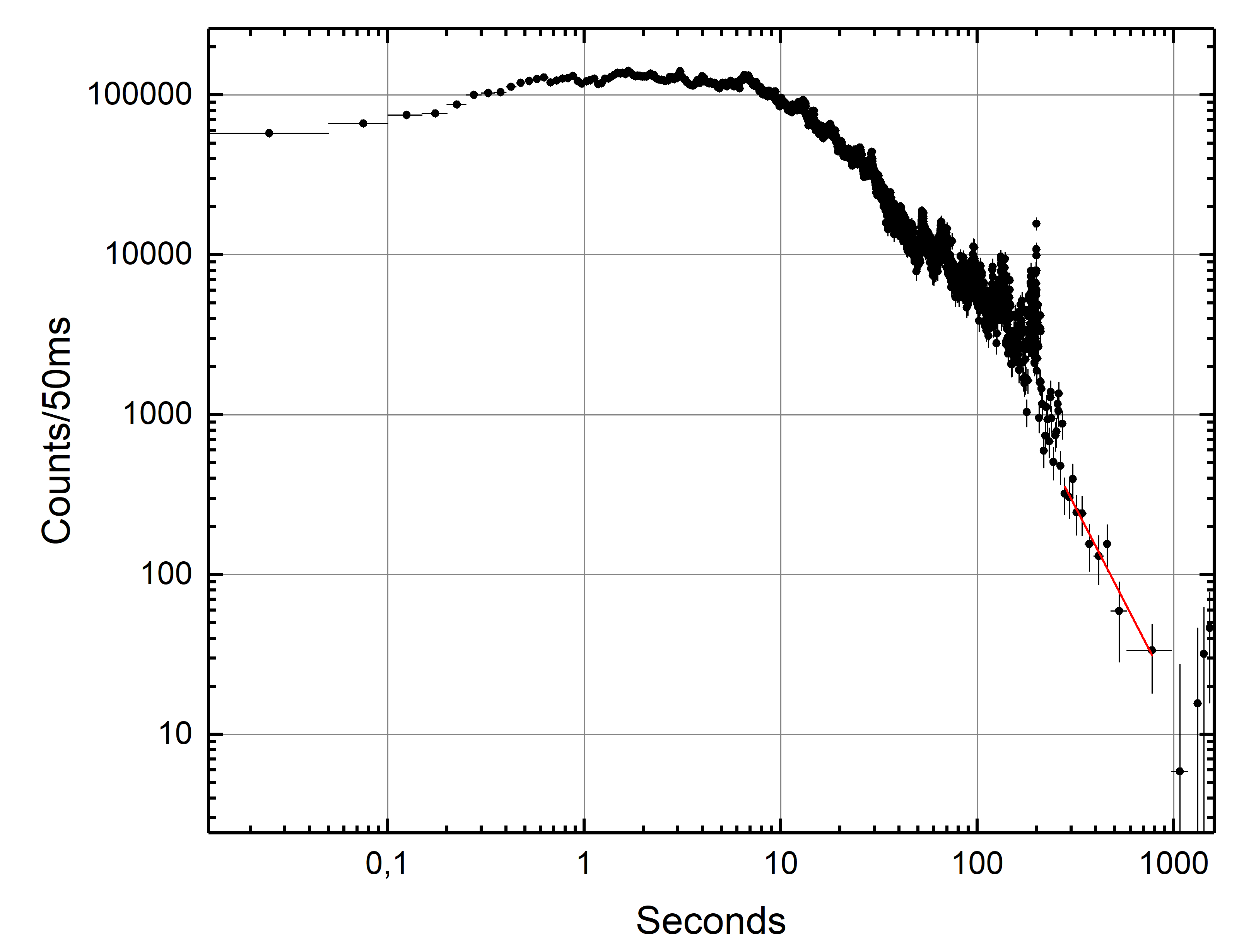}}
\caption{\rm Combined light curve of 308 long GRBs without extended emission in the individual light curves. The red line indicates the PL fit to the extended emission.}
\label{long_sum_log}
\end{figure}

The combined light curve of long bursts was con- structed in the interval $[-2000;2000]$ s. From 543 long $T_{90}>0.7$ s) bursts, without extended emission in their individual light curves, we selected 308 events in which $L,R\geq2000$ s. Figure~\ref{long_sum_log} presents the combined light curve. A significant signal up to $\sim800$ s is present in it, at a median $T_{90}$ of the sample $\sim25$ s. These values confirm the presence of extended emission in the combined light curve as well. The light curve can be fitted by a power law (Fig.~\ref{long_sum_log}, the red straight line) with an index $\alpha=2.4\pm0.3$.

\section*{RESULTS AND DISCUSSION}
\noindent

The duration parameter $T_{90}$ calculated here precisely for the bursts with the detected extended emission can considerably exceeed the values of $T_{90}$ from other catalogs. This is because the 95\% level of counts in the integrated light curve does include part of the interval in which the extended emission was found. At the same time, only the active phase falls into the $T_{50}$ interval even in the events with the most intense and longest extended emission (for example, GRB 021206 and GRB 130427). This suggests that, in contrast to $T_{90}$, $T_{50}$ can characterize better the duration of the active phase.

The $\frac{T_{90}}{T_{50}}$ distribution is presented in Fig.~\ref{t90_t50}. As expected, this ratio is systematically larger for bursts with extended emission (the corresponding median values are $\frac{T_{90}}{T_{50}}$ = 3 and $\frac{T_{90}}{T_{50}}$ = 5.6). All of the GRBs without extended emission have $\frac{T_{90}}{T_{50}}$ $< 80$. If, alternatively, $\frac{T_{90}}{T_{50}}>80$, then there is type I extended emission in the burst. However, this param- eter cannot be used as a criterion for the presence of extended emission. At $\frac{T_{90}}{T_{50}}<80$ the distributions overlap significantly and based only on this value, we cannot unambiguously determine the presence of extended emission. In this case, a more detailed study of the light curve is required.

Extended emission was found in 151 individual light curves of the investigated bursts, accounting for $\sim20\%$ of the total number of investigated events.

The light curves of the GRBs without extended emission in their individual light curves were combined (stacked) relative to the onset time of the gamma-ray emission. Extended emission was detected in the combined light curve of long bursts. A significant signal is observed up to 800 s, which is considerably larger than the median duration of the sample of the same bursts ($T_{90}\sim25$ s). Consequently, the detected signal may actually be deemed the extended emission. This confirms the results by Connaughton (2002), who analyzed the combined light curve of long bursts based on data from the BATSE experiment (20–100 keV), and the results by Burenin et al. (2000) based on SIGMA/Granat data (35–300 keV). The behavior of the light curve for the extended emission in the combined light curve (a power law decay with an index$\alpha=2.4\pm0.3$) suggests that it is probably associated with the superposition of light curves for the active phase at the decay stage and not with the additional component of the light curve, i.e., it is type II extended emission. Indeed, despite the fact that the detection rate of type II extended emission is lower than that of type I one by a factor of $\frac{105}{46}\sim2$, the mean intensity of this emission is higher by a factor of $\sim4$. Consequently, when a large number of light curves are combined, the type II extended emission will make a dominant contribution in the time range before $\sim 800$ seconds.

No extended emission was detected in the combined light curve of short GRBs. This results is in a contradiction with the result from Minaev et al. (2010a), where the extended emission was observed up to $\sim125$ s. A probable cause of the discrepancies is the smaller number of GRBs in this paper (40 versus 105 events).

\begin{figure}[h]
\hspace{-2cm}
\center{\includegraphics[width=1\linewidth]{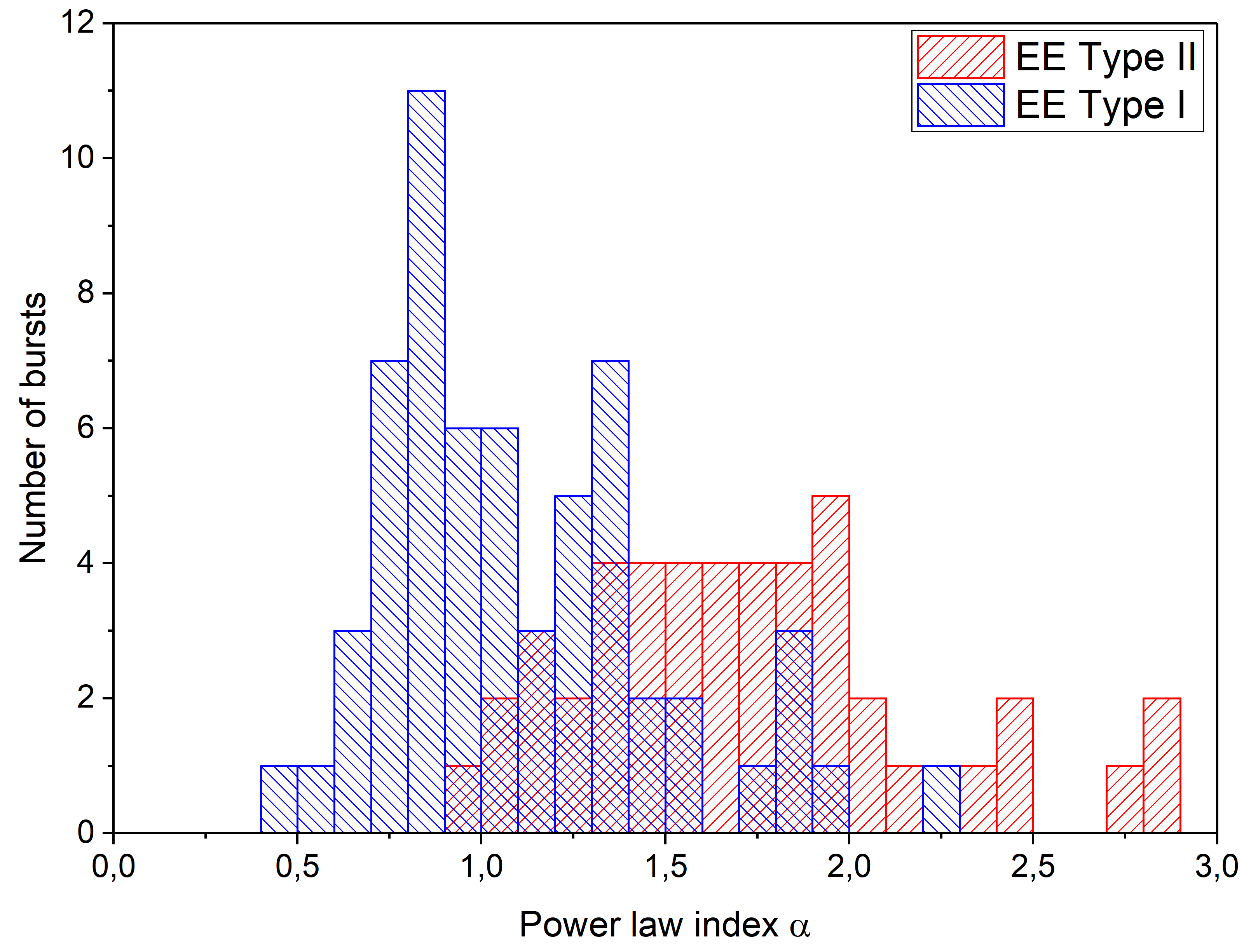}}
\caption{\rm Distribution of bursts in PL index $\alpha$ in the extended emission model. The blue and red colors mark the bursts with type I and II extended emission, respectively.}
\label{PLD}
\end{figure}

\begin{figure}[h]
\center{\includegraphics[width=1\linewidth]{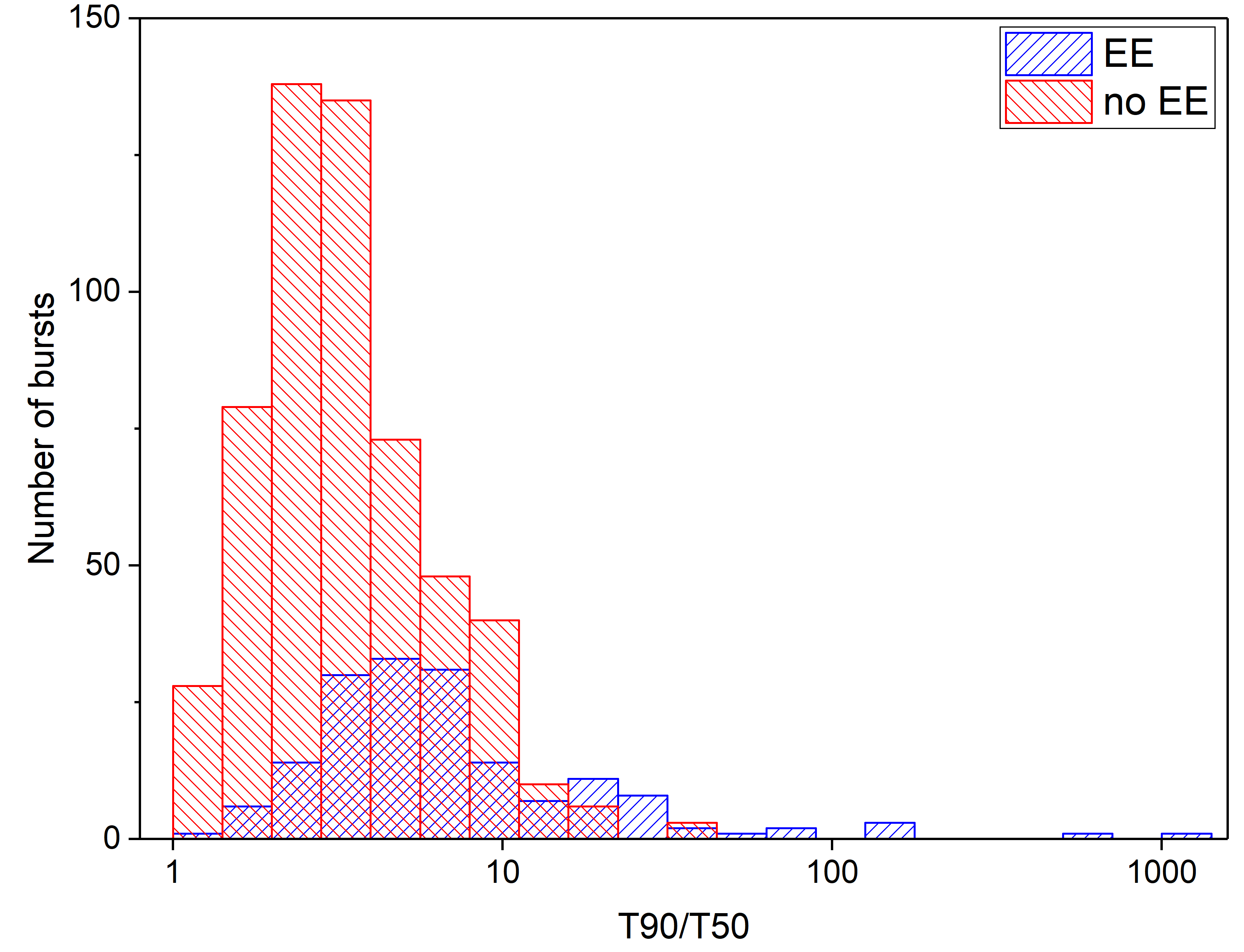}}
\caption{\rm Distribution of bursts in $\frac{T_{90}}{T_{50}}$. The blue color marks the bursts with a significant extended emission. The red color marks the bursts without it.}
\label{t90_t50}
\end{figure}

We analyzed the distribution of bursts in PL index $\alpha$ for the extended emission (Fig.~\ref{PLD}). The Kolmogorov–Smirnov test for two samples of extended emission showed that the probability that these distributions were obtained from single general population is $p=4\cross10^{-10}$,which confirms the existence of two phenomenologically different types of extended emission. The median values for groups I (105 events) and II (46 events) are  $\alpha$ = 1.0 and 1.8, respectively. A break between the prompt active phase and the extended emission is present in the light curve for the group I bursts, i.e., this emission is an additional component that is not associated with the active phase. The PL indices $\alpha$ are close to those of the light curve for the afterglow in the X-ray and optical bands. The type I extended emission is probably an early afterglow stage. The group II bursts are characterized by a larger PL index ($\alpha$ = 1.8), which is apparently the decay of the active phase.

\section*{CONCLUSIONS}
\noindent

We carried out a systematical search for extended gamma-ray ($>80$ keV) emission and its investigation for the brightest GRBs detected in the SPI- ACS/INTEGRAL experiment. Out of the 739 events, 45 belong to the class of short GRBs and the remaining 694 belong to the class of long ones. The fraction of bursts with extended emission is $\sim20\%$. The longest duration of the extended emission is 10 000 s. Our study of the extended emission became possible owing to the stable background in the orbit of the INTEGRAL space observatory.

Two types of extended emission characterized by a light curve with a PL decay and diﬀerent PL indices,(I) $\alpha$ = 1.0 and (II) $\alpha$ = 1.8, were detected in the individual light curves of long GRBs. The extended emission for the group I bursts is an additional com- ponent that is not associated with the active phase and is probably an early afterglow stage. The group II bursts are characterized by a larger PL index ($\alpha$ = 1.8), which is probably related to the continuation of the central engine operation and is the end of the active phase.

A statistically significant extended emission observed up to 800 s and fitted by a power law with an index $\alpha=2.4\pm0.3$ was also detected in the combined light curve of long bursts, for which no extended emission was found in the individual light curves (543 bursts). This confirms that the extended emission is a common property of all long bursts and is the sum of type I and II extended emissions.

We showed that the parameter $T_{50}$ better characterizes the duration of the prompt phase, while the parameter $T_{90}$ should be used with cation as a characteristic of the duration of the active phase.

\section*{ACKNOWLEDGEMENTS}
\noindent

This work was supported by RSF grant no. 18-12-00378.

\clearpage

\clearpage

\clearpage

\clearpage

\clearpage
\centering
\vspace{5mm}
\begin{longtable}[h]{l|c|c|c|c|c}
\captionsetup{font=bf}
\caption{Comparison of the durations $T_{90}$} \\
\hline\hline
GRB	&	$T_0$, UT	&	$T_{90}$\a, s	&	$\sigma_{T_{90}}^{-}$	&	$\sigma_{T_{90}}^{+}$	&	$T_{90}$\b, s		\\ \hline
\endfirsthead
021102	&	15:58:32	&	9.75	&	1.55	&	2.75	&	10.8[3]	\\
021116	&	08:06:34	&	45.9	&	10.7	&	12.5	&	-	\\
021125	&	05:59:01	&	19.9	&	1.9	&	3.4	&	-	\\
021125	&	17:58:27	&	31.8	&	6.1	&	15.7	&	67.5[3]	\\
021201	&	05:30:04	&	0.25	&	0	&	0.05	&	0.34[3]	\\
021206	&	22:49:11	&	2280	&	139	&	135	&	4.92[3]	\\
021226	&	14:53:40	&	0.8	&	0.2	&	2.4	&	0.35[3]	\\
021228	&	14:56:41	&	41.8	&	18.4	&	5.2	&	-	\\
030102	&	15:47:50	&	25.7	&	10.3	&	19.3	&	-	\\
030102	&	23:18:58	&	24.7	&	6.3	&	20.7	&	13.2[3]	\\
030105	&	14:34:14	&	14.4	&	5.9	&	3.9	&	1.23[3]	\\
030115	&	06:25:12	&	151	&	75	&	71	&	79.5[3]	\\
030117	&	17:36:14	&	0.2	&	0.1	&	0.05	&	-	\\
030127	&	12:32:32	&	64	&	14	&	37	&	38[3]	\\
030204	&	12:45:34	&	43.1	&	5.9	&	7.9	&	56[3]	\\
030215	&	11:16:22	&	35.8	&	0.4	&	0.3	&	-	\\
030217	&	23:31:42	&	0.35	&	0.1	&	0.3	&	-	\\
030218	&	11:42:38	&	183	&	2.6	&	3.2	&	-	\\
030220	&	16:12:44	&	98.2	&	5.2	&	4.9	&	-	\\
030223	&	09:45:06	&	38.6	&	11.8	&	5.7	&	20.5[3]	\\
030225	&	15:02:47	&	73.1	&	6.6	&	5.0	&	20[3]	\\
030307	&	14:31:58	&	3.5	&	0.35	&	0.41	&	3.8[3]	\\
030325	&	14:15:10	&	2.05	&	0.35	&	0.61	&	-	\\
030326	&	10:43:41	&	9.8	&	2.1	&	2.4	&	12.6[3]	\\
030329	&	11:37:15	&	23.8	&	1.4	&	1.7	&	17.4[3],21.808[4]	\\
030331	&	05:38:15	&	128.9	&	74.6	&	11.3	&	23.8[3]	\\
030406	&	22:42:03	&	116.7	&	14.6	&	32.1	&	70.2[3]	\\
030413	&	07:34:44	&	74.3	&	11.6	&	11.3	&	20.4[3]	\\
030414	&	13:48:27	&	24.8	&	3.7	&	1.8	&	28.5[3]	\\
030419	&	01:12:06	&	38.2	&	0.4	&	0.7	&	37.8[3]	\\
030501	&	01:17:17	&	18.5	&	3.0	&	7.1	&	7.4[3]	\\ \hline
\multicolumn{6}{l}{\a  $T_{90}$ derived from the SPI-ACS data in this paper.}\\
\multicolumn{6}{l}{\b $T_{90}$ derived from the data of other experiments.}\\
\multicolumn{6}{l}{$[1]$ $T_{90}$ from the GBM/Fermi catalog (50–300 keV) (Bhat et al. 2016).}\\
\multicolumn{6}{l}{$[2]$ $T_{90}$ from the Swift GRB table (15–150 keV).}\\
\multicolumn{6}{l}{$[3]$ $T_{90}$ from the RHESSI GRB table (25 keV–1.5 MeV) (Ripa et al. 2009).}\\
\multicolumn{6}{l}{$[4]$ $T_{90}$ from the Konus/Wind catalog ( 80 1200 keV) (Tsvetkova et al. 2017).}\\
\multicolumn{6}{l}{$[5]$ $T_{90}$ from the HETE GRB table (30–400 keV).}\\
\multicolumn{6}{l}{A complete version of the table is accessible in electronic form at}\\
\multicolumn{6}{l}{at http://vizier.u-strasbg.fr/viz-bin/VizieR}\\
\end{longtable}
\clearpage

\vspace{5mm}
\begin{longtable}{l|c|c|c|c|c|c|c}
\captionsetup{font=bf}
\caption{GRBs with a significant extended emission.} \\
\hline\hline
GRB	&	Best 	&	A\b					&	$\alpha$\c					&	$t_{EE}$\d					&	$\sigma$\e	&	$\frac{T_{90}}{T_{50}}$	&	Type	\\
 & Model\a & & & & & & \\\hline
\endfirsthead
021116	&	PL	&	$	4559	\pm	2445	$	&	$	1.53	\pm	0.17	$	&		-				&	8.8	&	3.6	&	IIa	\\
021206	&	biased PL	&	$	1280	\pm	104	$	&	$	0.76	\pm	0.01	$	&	$	8.15	\pm	0.3	$	&	27.3	&	692.0	&	Ia	\\
021228	&	PL	&	$	1001	\pm	211	$	&	$	1.38	\pm	0.11	$	&		-				&	6.5	&	5.3	&	IIa	\\
030105	&	-	&		-				&		-				&		-				&	6.4	&	28.8	&	Ia	\\
030218	&	PL	&	$	776	\pm	228	$	&	$	1.03	\pm	0.09	$	&		-				&	4.1	&	8.9	&	Ib	\\
030225	&	PL	&	$	7947	\pm	2045	$	&	$	1.46	\pm	0.08	$	&		-				&	16.3	&	3.1	&	IIa	\\
030326	&	PL	&	$	3108	\pm	403	$	&	$	1.92	\pm	0.10	$	&		-				&	4.2	&	3.8	&	IIa	\\
030331	&	-	&		-				&		-				&		-				&	3.4	&	24.8	&	Ia	\\
030413	&	PL	&	$	73	\pm	88	$	&	$	0.49	\pm	0.30	$	&		-				&	6.1	&	7.6	&	Ia	\\
030414	&	PL	&	$	19296	\pm	2522	$	&	$	1.90	\pm	0.06	$	&		-				&	38.4	&	4.9	&	IIa	\\
030506	&	-	&		-				&		-				&		-				&	6.3	&	4.4	&	Ia	\\
030717	&	PL	&	$	49	\pm	9	$	&	$	0.76	\pm	0.16	$	&		-				&	4.0	&	5.7	&	Ia	\\
030726	&	PL	&	$	16337	\pm	20641	$	&	$	1.47	\pm	0.28	$	&		-				&	11.6	&	4.4	&	Ia	\\
030801	&	PL	&	$	23080	\pm	8173	$	&	$	1.86	\pm	0.11	$	&		-				&	6.8	&	10.1	&	IIa	\\
030814	&	PL	&	$	1105	\pm	476	$	&	$	1.08	\pm	0.13	$	&		-				&	11.1	&	9.8	&	Ia	\\
030827	&	-	&		-				&		-				&		-				&	4.1	&	39.7	&	Ia	\\
031107	&	biased PL	&	$	1077	\pm	816	$	&	$	1.10	\pm	0.20	$	&	$	10.18	\pm	2.19	$	&	11.8	&	7.9	&	IIa	\\
031111	&	PL	&	$	74460	\pm	87521	$	&	$	2.80	\pm	0.50	$	&		-				&	6.7	&	5.5	&	IIa	\\
031127	&	PL	&	$	39901	\pm	14440	$	&	$	1.67	\pm	0.09	$	&		-				&	14.3	&	5.3	&	IIa	\\
031202	&	PL	&	$	102	\pm	54	$	&	$	0.50	\pm	0.13	$	&		-				&	9.0	&	17.6	&	IIa	\\
031214	&	PL	&	$	258	\pm	124	$	&	$	0.95	\pm	0.17	$	&		-				&	3.1	&	127.3	&	Ia	\\
031219	&	biased PL	&	$	2464	\pm	936	$	&	$	1.79	\pm	0.18	$	&	$	1.63	\pm	0.28	$	&	21.7	&	4.1	&	IIa	\\
040324	&	PL	&	$	120	\pm	12	$	&	$	0.96	\pm	0.10	$	&		-				&	5.2	&	15.0	&	Ia	\\
040421	&	-	&		-				&		-				&		-				&	5.3	&	1.7	&	Ia	\\
040425	&	PL	&	$	8983	\pm	259	$	&	$	1.80	\pm	0.02	$	&		-				&	39.7	&	3.3	&	IIa	\\
040612	&	-	&		-				&		-				&		-				&	5.3	&	6.0	&	Ia	\\
040615	&	PL	&	$	74032	\pm	16556	$	&	$	1.75	\pm	0.06	$	&		-				&	31.0	&	5.7	&	IIa	\\
040921	&	PL	&	$	1982	\pm	1019	$	&	$	1.23	\pm	0.17	$	&		-				&	13.4	&	6.8	&	Ia	\\
040922	&	PL	&	$	317731	\pm	70155	$	&	$	1.81	\pm	0.05	$	&		-				&	39.3	&	4.8	&	IIa	\\ \hline
\multicolumn{8}{l}{\a  The best model to fit the extended emission.}\\
\multicolumn{8}{l}{\b  The normalization parameter (amplitude) in the extended emission model.}\\
\multicolumn{8}{l}{\c  The PL index in the extended emission model.}\\
\multicolumn{8}{l}{\d  The parameter $t_{EE}$ in the biased PL extended emission model. If the best model is PL}\\
\multicolumn{8}{l}{   or the fit by PL and biased PL is unsatisfactory,}\\
\multicolumn{8}{l}{   then a dash is given.}\\
\multicolumn{8}{l}{\e The statistical significance of the extended emission.}\\
\multicolumn{8}{l}{Note. A complete version of the table is accessible in electronic form at}\\
\multicolumn{8}{l}{http://vizier.u-strasbg.fr/viz-bin/VizieR}\\
\end{longtable}
\clearpage



\end{document}